\documentclass[12pt]{article}

\usepackage{epsf,epsfig}
\usepackage{graphics}

\usepackage{cite}
\usepackage{amsmath,amssymb}
\def\slash#1{#1 \hskip-0.45em /}

\def\be{\begin{equation}}
\def\ee{\end{equation}}
\def\bea{\begin{eqnarray}}
\def\eea{\end{eqnarray}}

\def\nnb{\nonumber}
\def\eps{\epsilon}
\def\as{\alpha_s}

\def\np{n_+}
\def\nm{n_-}
\def\DB0{\partial B_0}

\def\Cl2{\mbox{Cl}_2}

\def\slash#1{#1 \hskip-0.45em /}

\newcommand{\PL}[2]{\, {\rm Li}_{#1}\!\left({#2}\right)}

\newcommand{\braket}[1]{\langle #1 \rangle}

\newcommand{\loopint}[1]{\int \!\!\! \frac{d^D #1}{\left(2\pi\right)^D}\!}
\newcommand{\ESGamma}{S_{\Gamma}}

\def\eps{\epsilon}

\def\MSbar{\overline{\rm MS}}

\begin{document}
\thispagestyle{empty}

\begin{flushright}
{\small
PITHA~08/25\\
SFB/CPP-08-80\\
0810.1230 [hep-ph]\\
October 7, 2008}
\end{flushright}

\vspace{\baselineskip}

\begin{center}
\vspace{0.5\baselineskip}
\textbf{\Large\boldmath
Two-loop QCD correction to \\[0.0cm] 
differential semi-leptonic $b \to u$ decays \\[0.2cm]
in the  shape-function region}\\
\vspace{3\baselineskip}
{\sc M.~Beneke}$^{a,b}$, {\sc T.~Huber}$^{a}$, and 
{\sc Xin-Qiang~Li}$^{a,}$\footnote{Alexander-von-Humboldt Fellow}\\
\vspace{0.7cm}
{\sl $^a$ 
Institut f\"ur Theoretische Physik E, 
RWTH Aachen University,\\
D--52056 Aachen, Germany}\\[0.2cm]
{\sl $^b$ 
Institut f\"ur Theoretische Physik, 
Universit\"at Z\"urich,\\
CH -- 8057 Z\"urich, Switzerland}\\[0.2cm]
\vspace{3\baselineskip}

\vspace*{0.2cm}
\textbf{Abstract}\\
\vspace{1\baselineskip}
\parbox{0.9\textwidth}{
We calculate the two-loop QCD correction to the form factors of 
on-shell $b$-quark decay to an energetic massless quark, which 
constitutes the last missing piece required for an 
${\cal O}(\alpha_s^2)$ determination of $|V_{ub}|$ from 
inclusive semi-leptonic $\bar B\to X_u\ell \bar \nu$ decays in the 
shape-function region. 
}
\end{center}

\newpage
\setcounter{page}{1}

\newpage
\allowdisplaybreaks[2]

\section{Introduction}
\label{sec:intro}

The strength of $b\to u$ transitions, measured by the CKM element 
$|V_{ub}|$, calibrates one of the sides of the unitarity triangle and 
is therefore an important input to flavour physics. Exclusive 
semi-leptonic $\bar B\to M \ell \bar \nu$ decays (with $M=\pi,\rho,\ldots$) 
provide direct access to the $b\to u$ transition, but 
require knowledge of the heavy-to-light meson form factors to 
extract $|V_{ub}|$. Inclusive semi-leptonic heavy quark decays on 
the other hand, can be calculated in perturbation theory. However, 
the need  to separate $\bar B\to X_u\ell \bar \nu$ from an overwhelming 
charm final-state background requires cuts on the differential 
decay spectra, which in one way or another constrain the hadronic 
final state to have small invariant mass and large energy. The final 
state distribution then depends on a non-perturbative function describing 
the light-cone residual momentum distribution of the $b$-quark in 
the $\bar B$ meson, the shape function \cite{Neubert:1993ch,Bigi:1993ex}. 
Somewhat unfortunately, the two methods currently result in values of 
$|V_{ub}|$ that seem to systematically differ, with the exclusive 
method favouring smaller $|V_{ub}|$. This provides motivation and urgency 
to improving the theoretical accuracy of both methods.

The differential distributions in inclusive semi-leptonic decay 
are calculated from the discontinuity 
\begin{equation}
W^{\mu\nu} = \frac{1}{\pi} \,\mbox{Im}\left[\langle\bar B(p_B)|
\,i\!\int d^4 x \,e^{i q \cdot x} \,T(j^{\dagger\mu}(0) j^\nu(x))
|\bar B(p_B)\rangle\right]
\end{equation}
of the forward matrix element of a correlation function of the 
weak interaction current $j^\mu = \bar u\gamma^\mu(1-\gamma_5)b$.
In the shape-function region the structure functions into which 
the hadronic tensor $W^{\mu\nu}$ can be decomposed, can be organized 
into the factorized expression~\cite{Korchemsky:1994jb}
\begin{equation}
W_k = \sum_{i,j} \,a^{ij}_k \,C_i(n_-\cdot p) \,C_j(n_-\cdot p)
\int d\omega \,J(p_\omega^2)\,S(\omega)
\label{wi}
\end{equation}
at leading power in the $1/m_b$ expansion. The factors 
$C_i C_j$, $J$, $S$ arise from the different scales $m_b$, 
$\sqrt{m_b\Lambda_{\rm QCD}}$ and $\Lambda_{\rm QCD}$, respectively, 
such that $C_i$ and $J$ can be computed in QCD perturbation theory, while 
the $B$ meson shape function is non-perturbative. ($a^{ij}_k$ 
are numerical constants.) Eq.~(\ref{wi}) 
has been worked out at order ${\cal O}(\alpha_s)$ 
in~\cite{Bauer:2003pi,Bosch:2004th} and 
forms the basis of the inclusive $|V_{ub}|$ analysis 
performed in \cite{Lange:2005yw}. A summary of other inclusive $|V_{ub}|$ 
analyses and recent experimental results based on these methods 
is given in \cite{Barberio:2008fa}.
At the two-loop order, the jet function $J$ and 
the partonic shape function of the $b$ quark have already been 
calculated~\cite{Becher:2005pd,Becher:2006qw}. Here we
compute the two-loop hard matching coefficients $C_i$ associated 
with the form factors of on-shell $b$-quark decay into 
an energetic massless quark and provide a numerical estimate of 
the new term. All items for a ${\cal O}(\alpha_s^2)$ determination 
of $|V_{ub}|$ from inclusive semi-leptonic decay are now in place, 
which should remove most of the perturbative theoretical uncertainty. 
A detailed phenomenological analysis is, however, beyond the scope 
of the current work. 

While this paper was in preparation, the calculation reported here 
has also been completed by Bonciani and 
Ferroglia~\cite{Bonciani:2008wf}. The results have been compared 
prior to publication and complete agreement has been found. On the day 
of submission of the present work, 
the paper \cite{Asatrian:2008uk} on the same 
topic has appeared.
 
\section{Structure of the calculation}
\label{sec:structure}

\subsection{Set-up of the matching calculation}
\label{subsec:notation}

The calculation of the short-distance coefficients $C_i$ in 
(\ref{wi}) amounts to 
matching the current $\bar u\gamma^\mu(1-\gamma_5)b$ to a set 
of leading-power currents in soft-collinear effective theory 
(SCET)~\cite{Bauer:2000yr,Bauer:2001yt,Beneke:2002ph}. One-loop 
results for the $C_i$ have been obtained in this framework 
in~\cite{Bauer:2000yr,Beneke:2004rc}. The current is represented 
in SCET by 
\begin{equation}
[\bar u\gamma^\mu(1-\gamma_5)b](0) = 
\int d\hat{s} \sum\limits_{i} \widetilde{C}_{i}(\hat{s})\,
(\bar{\xi} W_c)(sn_+) \Gamma_i^\mu h_v(0) + \ldots,
\label{matchdef}
\end{equation}
where $\xi$ is the collinear up-quark field in SCET, $h_v$ the 
static heavy-quark field, and $W_c$ a collinear Wilson line. 
(We use notation as defined in more detail in \cite{Beneke:2004rc}.) 
There are three independent Dirac matrices that can appear 
on the right-hand side, which we choose as
\be
\label{eq:diracbasis}
\Gamma_1^\mu = \gamma^{\mu} (1-\gamma_5) \; ,\qquad 
\Gamma_2^\mu = v^{\mu} (1+\gamma_5) \; ,\qquad 
\Gamma_3^\mu = \nm^{\mu} (1+\gamma_5)
\; .
\ee
The ellipses in (\ref{matchdef}) 
denote higher-dimensional operators, which are not 
relevant to the present work, and $\hat{s}\equiv s m_b$. SCET 
makes extensive use of two light-like vectors $n_+^\mu$, $n_-^\mu$ 
with $n_+\cdot n_-=2$, with respect to which four-vectors are 
decomposed as
\be
p^{\mu}  = \np \cdot p \, \frac{\nm^\mu}{2}+ 
\nm \cdot p \,  \frac{\np^\mu}{2} + p_{\perp}^\mu ,
\ee
with $\nm \cdot p_\perp = \np \cdot p_\perp=0$. The collinear field 
$\xi$ describes modes which have $n_+\cdot p$ large, of order $m_b$. 
To define the heavy-quark field, we choose a frame such that 
$v^\mu=(\np^{\mu}+\nm^{\mu})/2$, that is $n_+\cdot v=n_-\cdot v=1$.
The factorization formula (\ref{wi}) for the differential decay 
distributions makes use of the momentum-space short-distance coefficient 
rather than the position-space expression appearing in the convolution 
in (\ref{matchdef}). The momentum space
coefficient functions are related to those defined above by
\begin{equation}
  C_i(u)=\int d\hat{s} \, 
  e^{i u \hat{s}}\,\widetilde{C}_i(\hat{s}),
\label{ft}
\end{equation}
where the new variable $u\in [0,1]$ equals the momentum fraction 
$n_+ \cdot p/m_b$ of the external up-quark line
in a momentum-space Feynman diagram.

The actual matching calculation is also done in momentum space and
yields the momentum-space coefficient functions directly. To this end 
we consider the matrix element of the current between a 
bottom quark of mass $m_b$ and momentum $p_b$ and a massless 
up-quark with momentum $p$. Both quarks are considered on-shell, hence
$p_b^2=m_b^2$ and $p^2=0$.  The absence of any perturbative infrared 
scale implies that the SCET loop diagrams are scaleless, and 
the SCET matrix element is given by its tree-level expression multiplied 
by the universal renormalization factor for the SCET currents 
$(\bar{\xi} W_c)\Gamma_i^\mu h_v$: 
\begin{equation}
\braket{u(p)|(\bar{\xi} W_c)(sn_+) \Gamma_i^\mu h_v(0)|b(p_b)} 
=  e^{i s n_+\cdot p} \,Z_J\,\bar u_{n_-} \, \Gamma_i^\mu \, u_v \, .
\label{scetme1}
\end{equation}
Thus, calculating the QCD matrix element in dimensional 
regularization yields directly 
the dimensionally regularized short-distance coefficient $C_i(u)$.
To this end, we first decompose the matrix element of the 
hadronic current into three form factors according to
\bea
\braket{u(p)|\bar u\gamma^{\mu} (1-\gamma_5) b|b(p_b)} &=& 
F_1(u) \; \bar u(p) \gamma^{\mu} (1-\gamma_5) u(p_b) + 
F_2(u) \; \bar u(p)
\frac{p_b^{\mu}}{m_b} (1+\gamma_5) u(p_b) \nnb \\
&& + F_3(u) \; \bar u(p)\frac{m_b \, p^{\mu}}
{p_b \cdot p} (1+\gamma_5) u(p_b) \; .
\label{eq:MEQCD}
\eea
The form factors $F_i(u)$, $i=1,2,3$ can only depend on the dimensionless
variable
$u \equiv 2 \, p_b \cdot p/m_b^2$
and logarithms of $\mu^2/m_b^2$, where $\mu$ is at once 
the renormalization scale of the strong coupling and the infrared 
factorization scale, since the on-shell heavy-to-light form factors 
contain soft and collinear divergences. Only $F_1$ is
 non-zero at tree-level, $F_1 = 1 + {\cal
O}(\alpha_s)$, while $F_{2,3} = {\cal O}(\alpha_s)$. Identifying 
$p_b^\mu=m_b v^\mu$ and $n_-^\mu=m_b p^\mu/(p_b\cdot p)$, we 
see that $u$ equals the variable $u$ defined in (\ref{ft}). At leading 
order in the heavy-quark expansion the spinors of the quark fields 
equal $\bar u(p) = \bar u_{n_-}$, $u(p_b) = u_v$, 
where the collinear and heavy quark spinors satisfy $\slash{n}_-u_{n_-}=0$ 
and $\slash{v}u_v=u_v$, respectively. Eq.~(\ref{eq:MEQCD}) then becomes
\bea
\braket{u(p)|\bar u\gamma^{\mu} (1-\gamma_5) b|b(p_b)} &=& 
\sum_{i=1}^3 F_i(u) \; \bar u_{n_-} \, \Gamma_i^\mu \, u_v ,
\label{eq:MESCET}
\eea
while using (\ref{scetme1}) the matrix element of the right-hand side 
of (\ref{matchdef}) 
equals
\begin{equation}
\sum_{i=1}^3 Z_J\, C_i(n_+\cdot p/m_b)\; \bar u_{n_-} \, \Gamma_i^\mu \, u_v,
\end{equation}
so we simply have $C_i(u) = Z_J^{-1} F_i(u)$, $i=1,2,3$. In the following we 
briefly describe the method of calculating the two-loop QCD correction to 
the $F_i$.

\subsection{Calculational methods}
\label{subsec:method}

\begin{figure}[t]
\centering
\includegraphics[width=0.75\textwidth]{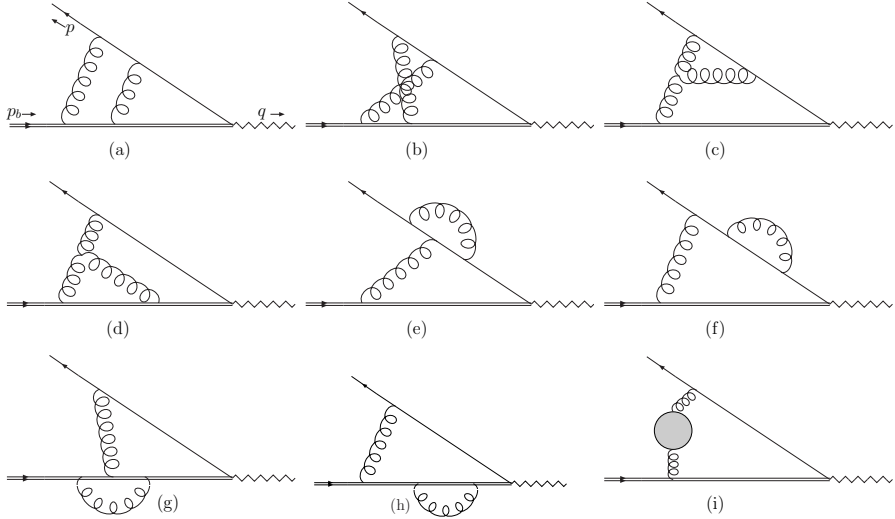}
\caption{Two-loop diagrams needed for the calculation. Double straight lines stand for massive quarks of mass $m_b$, whereas
single ones stand for massless quarks. The filled circle in the last diagram represents the complete one-loop gluon self-energy with
$n_f$ massless quarks (including the charm) and the massive bottom quark. \label{fig:2loopdiag}}
\end{figure}

The computation of the two-loop QCD vertex corrections to 
semi-leptonic $b \to u$ decays involves the evaluation of the diagrams shown
in Fig.~\ref{fig:2loopdiag}. We work in dimensional regularization 
with $D=4-2\eps$, where UV and IR (soft and collinear) divergences
appear as poles of up to the fourth order in $\eps$. For the matrix 
$\gamma_5$ we adopt the naive dimensional regularization (NDR) scheme
with anticommuting $\gamma_5$. 
Furthermore, we treat all quarks except the bottom quark as massless. We
checked that it is possible to include the effects of a non-zero 
charm mass analytically. However, since in the two-loop
calculation of the jet and shape function~\cite{Becher:2005pd,Becher:2006qw} 
the charm mass is neglected, we also set it to zero in the present work.

The amplitude of the diagrams is reduced by techniques that have become 
standard in multi-loop calculations. We apply a 
Passarino~--~Veltman~\cite{Passarino:1978jh} reduction of the  vector and 
tensor integrals. The Dirac and color algebra is then performed by
means of an in-house Mathematica routine. The dimensionally 
regularized scalar integrals are further reduced to a small set of 
master integrals (depicted in Fig.~\ref{fig:masters} in the Appendix) 
using the Laporta
algorithm~\cite{Laporta:1996mq,Laporta:2001dd} based on 
integration-by-parts~(IBP) 
identities~\cite{Tkachov:1981wb,Chetyrkin:1981qh}. To this end we use 
the Maple package AIR~\cite{Anastasiou:2004vj}. 

The techniques we apply during the evaluation of the master 
integrals are manifold. The easier integrals can be written in a 
closed form in
terms of $\Gamma$-functions and hypergeometric functions and subsequently expanded in $\eps$ with the package {\tt
HypExp}~\cite{Huber:2005yg,Huber:2007dx}. In more complicated cases we derive Mellin-Barnes representations by means of the package
AMBRE~\cite{Gluza:2007rt}. We perform the analytic
continuation to $\eps=0$ with the package MB~\cite{Czakon:2005rk}, which is also used for numerical cross-checks. We then apply Barnes'
lemmas and the theorem of residues on the multiple Mellin-Barnes integrals, and insert integral representations of hypergeometric
functions as well as $\psi$-functions and Euler's $B$-function where appropriate. As a third technique we apply the method of
differential equations~\cite{Kotikov:1990kg,Kotikov:1991hm,Kotikov:1991pm} and evaluate the boundary condition with the Mellin-Barnes
technique. This renders, for instance, a three-dimensional MB representation in the case of the crossed six-line master integral of
Fig.~\ref{fig:masters}(g) at $u=1$. Eventually, the master integrals 
are evaluated as Laurent series in $\eps$, with expansion coefficients 
of argument $u$ expressed analytically in logarithms and polylogarithms 
of increasing weight. The maximum weight which appears
in our calculation is four, and we can express all functions but 
one in terms of ordinary polylogarithms. The function that cannot be
expressed in terms of ordinary polylogarithms is the harmonic polylogarithm~\cite{Remiddi:1999ew} ${\rm HPL}(\{-2,2\},1-u)$. We left,
however, also a second function, ${\rm HPL}(\{-1,2\},1-u)$, in the HPL notation since its expression in terms of ordinary polylogarithms
is rather complicated~\cite{Maitre:2005uu}. The master integrals 
have been calculated already in~\cite{Bell:2006tz} and used in~\cite{Bell:2007tv}. We find (almost) perfect agreement on the
expressions in ~\cite{Bell:2006tz}. Moreover, we improve the numerical accuracy of one of the boundary conditions, see
Appendix~\ref{ap:masters}.

\section{Renormalization}
\label{sec:renormalization}

After the pure two-loop calculation the result still contains 
divergences of UV as well as IR (soft and collinear) nature. The former
divergences are cancelled after addition of the UV counterterms, 
the latter disappear after inclusion of the jet and shape function
contributions, since the partonic differential decay distributions 
are infrared-finite. In the following we also present the pole parts 
of the form factors $F_i$. Subtracting the IR poles as discussed 
in Sec.~\ref{sec:conclude} leads to the  
$\overline{\rm MS}$ definition of the SCET current.

\subsection{UV renormalization}
\label{subsec:UV}
In performing the UV renormalization we adopt the on-shell scheme for the heavy quark mass as well as for the heavy and light quark
field. The strong coupling $\as$ on the other hand is renormalized in the $\MSbar$ scheme. The respective renormalization constants
read~\cite{Broadhurst:1991fy,Gray:1990yh,Melnikov:2000zc}
\bea
Z_m^{\rm os} &=& 1 - C_F \, \frac{g_0^2 (m^2)^{\frac{D-4}{2}}}{(4\pi)^{D/2}}  \, \frac{ (D-1) \Gamma(2-\frac{D}{2})}{(D-3)} \; , \\
Z_h^{\rm os} &=& 1 - C_F \, \frac{g_0^2 (m^2)^{\frac{D-4}{2}}}{(4\pi)^{D/2}}  \, \frac{ (D-1) \Gamma(2-\frac{D}{2})}{(D-3)} \nnb \\
             && + \frac{g_0^4 (m^2)^{D-4}}{(4\pi)^{D}} \, \Gamma^2(3-\frac{D}{2}) \left\{C_F^2\left[\frac{18}{(D-4)^2}-\frac{51}{2(D-4)}\right.\right.\nnb\\
	     &&\left. +\frac{433}{8}-13\pi^2+16\pi^2\ln(2)-24\zeta_3 + {\cal O}(D-4)\right]+ C_F \, C_A
	        \left[-\frac{22}{(D-4)^2}\right. \nnb \\
	     && \left.+\frac{101}{2(D-4)} -\frac{803}{8}+5\pi^2-8\pi^2\ln(2)+12\zeta_3 + {\cal
	     O}(D-4)\right]\nnb\\
	     && + C_F \, n_f \, t_f \left[\frac{8}{(D-4)^2}-\frac{18}{(D-4)}+\frac{4\pi^2}{3}+\frac{59}{2}+ {\cal O}(D-4)\right]\nnb \\
	     && \left.+ C_F \, t_f \left[\frac{16}{(D-4)^2}-\frac{38}{3(D-4)}-\frac{16\pi^2}{3}+\frac{1139}{18}+ {\cal O}(D-4)\right] \right\}\; .
\eea
Here $g_0$ is the bare QCD coupling. $C_F = (N^2-1)/(2\,N)$ and $C_A=N$ are the Casimir operators of the fundamental and adjoint
representation of SU$(N)$, respectively, and $t_f=1/2$ denotes the normalization of the trace of two fundamental generators. $n_f$
stands for the number of massless quarks, and we set the number of heavy quarks of mass $m$ to unity throughout the paper.
The renormalization constant for a massless quark field in the on-shell scheme receives corrections only from two-loop and higher due to diagrams with 
massive quark loops. At two loops only a single diagram contributes 
and the $Z$ factor assumes the following closed form
\be
Z_l^{\rm os} = 1 + 2 \, C_F \, t_f \, \frac{g_0^4 (m^2)^{D-4}}{(4\pi)^D}  \, \frac{ (D-1)
\Gamma(4-\frac{D}{2})\Gamma(-\frac{D}{2})}{(D-5) (D-7)} \; .
\ee
The renormalization constant $Z_{\alpha}$ of the strong coupling in the $\MSbar$ scheme reads
\be
Z_{\alpha} = 1 + \frac{\alpha_s}{4\pi} \left[\frac{11}{3} C_A - \frac{4}{3} \, t_f \left(n_f+1\right)\right] \frac{2}{(D-4)} + {\cal
O}(\as^2)\; .
\label{zalpha}
\ee

The renormalization of the quark fields
amounts to mere multiplications. Expanding the bare and renormalized amplitude as well as the renormalization constants according
to
\bea
{\cal A}_{\rm{bare}} &=& {\cal A}^{(0)} + g_0^2 \, {\cal A}_{\rm{bare}}^{(1)}+ g_0^4 \, {\cal A}_{\rm{bare}}^{(2)} +{\cal O}(g_0^6)
\; , \\
{\cal A}_{\rm{ren}} &=& {\cal A}^{(0)} + g_0^2 \, {\cal A}_{\rm{ren}}^{(1)}+ g_0^4 \, {\cal A}_{\rm{ren}}^{(2)} +{\cal O}(g_0^6)
\; , \\
Z_i &=& 1 + g_0^2 \, \delta Z_i^{(1)} + g_0^4 \, \delta Z_i^{(2)} +{\cal O}(g_0^6)
\; , \qquad i = m,h,l, 
\eea
we find for the renormalized amplitude
\bea
{\cal A}_{\rm{ren}}^{(1)} &=&{\cal A}_{\rm{bare}}^{(1)} +\frac{1}{2} \, \delta Z_h^{(1)} \, {\cal A}^{(0)} \; , \\
{\cal A}_{\rm{ren}}^{(2)} &=&{\cal A}_{\rm{bare}}^{(2)} + \frac{1}{2} \, \delta Z_h^{(1)}\, {\cal A}_{\rm{bare}}^{(1)}+
\left[\frac{1}{2} \, \delta Z_h^{(2)}-\frac{1}{8} \, (\delta Z_h^{(1)})^2+\frac{1}{2} \, \delta Z_l^{(2)}\right] \, {\cal A}^{(0)}\; .
\label{eq:twoloopren}
\eea
The bare coupling $\alpha_s^0=g_0^2/(4\pi)$ is then renormalized simply 
by the procedure $\alpha_s^0 = Z_\alpha \, \alpha_s \,
\tilde\mu^{2 \epsilon}$ with $\tilde\mu^2 = \mu^2 \, 
{\rm exp}(\gamma_E-\ln4\pi)$. This can also be seen from the way we 
present our results in~(\ref{eq:Ci}), where $Z_\alpha$ accounts for the 
renormalization of the coupling constant.

\begin{figure}[t]
\centering
\includegraphics[width=0.35\textwidth]{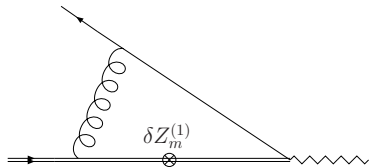}
\caption{One-loop counterterm diagram for mass renormalization \label{fig:massren}}
\end{figure}

The only non-trivial contribution to the UV renormalization is therefore the one-loop diagram in Fig.~\ref{fig:massren}. To this end,
only that part of the counterterm Feynman rule that contains the one-loop correction $\delta Z_m^{(1)}$ to $Z_m$ has to be inserted,
and the contribution has to be added to the RHS of Eq.~(\ref{eq:twoloopren}).
\subsection{IR subtraction, jet and shape 
function contribution}
\label{subsec:IR}

Since the partonic structure functions $W_i$ in (\ref{wi}) are 
infrared-finite, the $1/\epsilon$ poles of the two-loop 
coefficients $C_i$ or, equivalently, $F_i$ can be constructed 
independently using the two-loop expressions of the jet and shape 
functions~\cite{Becher:2005pd,Becher:2006qw} and the 
one-loop coefficients $F_i$ including their ${\cal O}(\epsilon)$ 
parts (given below). To check our calculation, we perform the convolution 
$\int d\omega \,J(p_\omega^2)\,S(\omega)$ of the 
unrenormalized jet and shape function with ${\cal O}(\alpha_s^2)$ 
accuracy and determine the infrared pole part of each $F_i$ individually 
by forming appropriate combinations of $i,j$. We find complete 
agreement with the poles obtained in the direct two-loop calculation 
of the $F_i$. 

In order to verify the pole cancellation in (\ref{wi}), 
one must remember that the calculation of the short-distance 
coefficients $C_i$ is performed 
in a theory with five active flavours (the fifth being the bottom 
quark), hence $\alpha_s$ that appears above is $\alpha_s^{(5)}$, 
while the jet and shape function are computed in four-flavour 
SCET. In combining $C_i$, $J$ and $S$ according to (\ref{wi}), 
the perturbative expansion of $C_i$ must be expressed in terms 
of  $\alpha_s^{(4)}$. The $D$-dimensional relation between 
the renormalized couplings required for 
this purpose reads to one-loop accuracy
\be\label{eq:asD}
\alpha_s^{(5)} =  \xi_{45} \, \alpha_s^{(4)} \, , \quad {\rm{where}} \quad \xi_{45} = 1+ 
\frac{{\alpha_s^{(4)}}}{4\pi} \, \frac{4 \, t_f}{3} \!\left[
e^{\gamma_E\epsilon}\,\Gamma(\epsilon) \left(\frac{m_b^2}{\mu^2}
\right)^{\!-\epsilon}-\frac{1}{\epsilon}\right]
\ee
and the couplings are evaluated at the scale $\mu$.

The $\overline{\rm MS}$ renormalized coefficient functions are obtained 
from $C_i = Z_J^{-1} F_i$ (see above), where \cite{Bauer:2000yr}
\begin{equation}\label{eq:ZJ}
Z_J = 1 +\frac{\alpha_s^{(4)} C_F}{4\pi} \left\{-\frac{1}{\epsilon^2} + 
\frac{1}{\epsilon}\left[-\ln\left(\frac{\mu^2}{u^2 m_b^2}\right)-
\frac{5}{2}\right]\right\} + {\cal O}(\alpha_s^2)
\end{equation}
at one loop, and the two-loop renormalization factor can be 
obtained from the result of our calculation by requiring that 
$C_1$ is free of $1/\epsilon$ poles at ${\cal O}(\alpha_s^2)$.

\section{Results}
\label{sec:results}

We present the result for the short-distance coefficients (form factors) 
in the form
\be\label{eq:Ci}
F_i = F_i^{(0)} + \frac{Z_{\alpha}  \as}{4\pi} \, F_i^{(1)} +\frac{Z^2_{\alpha} \, \as^2}{(4\pi)^2} \, F_i^{(2)}  + {\cal O}(\as^3) \; ,
\ee
with
\be
F_1^{(0)} = 1 \, , \qquad F_2^{(0)} = 0\, , \qquad F_3^{(0)} = 0 \; .
\ee
The factor $Z_\alpha$ is given in (\ref{zalpha}), and $Z_\alpha$ as well
as $\alpha_s$ in~(\ref{eq:Ci}) refer to a theory with five 
active quark flavors, $n_f=4$ massless ones and one massive one of 
mass $m_b$. Note that we expressed the series in terms of 
$Z_{\alpha}  \as$ rather than the renormalized coupling $\alpha_s$. 
Thus, the $F_i^{(2)}$ contain $1/\epsilon$ poles that are cancelled 
by charge renormalization as well as the IR poles discussed above.

\subsection{One-loop results}
\label{subsec:oneloop}

The one-loop hard matching coefficients are available in the literature 
through order 
${\cal O}\left(\eps^0\right)$~\cite{Bauer:2000yr,Beneke:2004rc,Bosch:2004th}, 
but the two-loop analysis of the structure functions $W_i$ requires 
them up to order ${\cal O}\left(\eps^2\right)$. Here we summarize the 
corresponding expressions for the form factors.
\bea
F_1^{(1)} &=& C_F \left\{-\frac{1}{\eps^2}+\frac{1}{\eps}\left[f_{-1}(u) - L\right]
-\frac{L^2}{2}+L \, f_{-1}(u) + f_{0}(u)\right. \nnb \\
&&+\eps\left[-\frac{L^3}{6}+ \frac{1}{2} \, L^2 \, f_{-1}(u)+L \, f_0(u) + f_1(u)\right]\nnb \\
&&\left.+\eps^2\left[-\frac{L^4}{24} + \frac{1}{6} \, L^3 \, f_{-1}(u)+ \frac{1}{2} L^2 \, f_{0}(u) + L \, f_1(u) +f_2(u)\right] \right\}\;,
\eea
\bea
F_2^{(1)} &=& C_F \left\{g_0(u) +\eps\left[ g_1(u) + L \, g_0(u)\right]
          +\eps^2\left[ g_2(u) + L \, g_1(u) +\frac{L^2}{2} \, g_0(u)\right]   \right\} \; ,\\
F_3^{(1)} &=&  C_F \left\{h_0(u) +\eps\left[ h_1(u) + L \, h_0(u)\right]
          +\eps^2\left[ h_2(u) + L \, h_1(u) +\frac{L^2}{2} \, h_0(u)\right]   \right\} \, ,
\eea
with
\be
L \equiv \ln\left(\frac{\mu^2}{m_b^2}\right) \; ,
\ee
and
\bea
f_{-1}(u) &=& 2\ln(u) - \frac{5}{2} \; , \\
f_{0}(u)  &=& -2 \ln^2(u)+2 \ln(u)\ln(1-u)+2 \PL{2}{u} + \frac{\ln(u)}{u-1}+3 \ln(u) -\frac{5\pi^2}{12}-6 \; ,\\
f_{1}(u)  &=& \frac{4}{3} \ln^3(u) -2 \ln^2(u)\ln(1-u) - \frac{\ln^2(u)}{u-1}-3 \ln^2(u)+ \frac{\ln(1-u)\ln(u)}{u-1} \nnb \\
          && +3 \ln(u)\ln(1-u)+ \frac{4\ln(u)}{u-1} +\frac{5\pi^2}{6} \, \ln(u) + 8 \ln(u) + \frac{\PL{2}{u}}{u-1}+ 3 \PL{2}{u}\nnb \\
	  && - 2 \PL{3}{1-u}- 4 \PL{3}{u} - \frac{\pi^2}{6 \, (u-1)} + \frac{13}{3} \, \zeta_3 -\frac{17\pi^2}{24} -12\; ,\\ 
f_{2}(u)  &=& -\frac{1}{6} \ln^4(1-u) - \frac{2}{3} \ln^4(u) + \frac{4}{3} \ln^3(u)\ln(1-u) + \frac{2 \ln^3(u)}{3\,(u-1)} + 2 \ln^3(u)\nnb \\
          && -\frac{\pi^2}{3} \ln^2(1-u) - \ln^2(u)\ln^2(1-u) - \frac{\ln(1-u)\ln^2(u)}{u-1} - 3 \ln(1-u)\ln^2(u) \nnb \\
	  &&- \frac{4\ln^2(u)}{u-1} - \frac{5\pi^2}{6} \ln^2(u) - 8 \ln^2(u) + \frac{2}{3} \ln(u)\ln^3(1-u)+ \frac{4\ln(1-u)\ln(u)}{u-1}\nnb \\
	  && + \frac{5\pi^2}{6} \ln(u)\ln(1-u) + 8 \ln(u)\ln(1-u) + \frac{5\pi^2\ln(u)}{12\,(u-1)} + \frac{8\ln(u)}{u-1} + \frac{5\pi^2}{4} \ln(u)\nnb \\
	  && + 16 \ln(u) + \frac{4 \PL{2}{u}}{u-1} + \frac{\pi^2}{6} \PL{2}{u}+ 8 \PL{2}{u} - \frac{\PL{3}{1-u}}{u-1}- 3 \PL{3}{1-u}\nnb \\
	  && - \frac{2\PL{3}{u}}{u-1}- 6 \PL{3}{u} + 2 \PL{4}{1-u} + 4 \PL{4}{u} -4 \PL{4}{\frac{u}{u-1}} - \frac{2\pi^2}{3\,(u-1)}\nnb \\
	  && - \frac{14}{3} \ln(u) \, \zeta_3 + \frac{2\zeta_3}{u-1} +\frac{41}{6} \, \zeta_3 - \frac{5\pi^4}{32}-
	  \frac{11\pi^2}{6}-24\; , \\
g_{0}(u)  &=& \frac{2 \, u \ln(u)}{(u-1)^2}-\frac{2}{u-1}\; , \\
g_{1}(u)  &=& - \frac{2 \, u \ln^2(u)}{(u-1)^2} + \frac{2 \, u \ln(u)}{(u-1)^2}- \frac{2 \, u \PL{2}{1-u}}{(u-1)^2} - \frac{4}{u-1}\; , \\
g_{2}(u)  &=& \frac{2 \, u }{(u-1)^2} \left[\ln(u)\ln(1-u)+\frac{2}{3} \ln^3(u)-\ln^2(u)\ln(1-u)-\ln^2(u)+\frac{5 \, \pi^2}{12}\ln(u)\right. \nnb\\
	  && \left. + 2 \ln(u)+ \PL{2}{u} - \PL{3}{1-u}- 2 \PL{3}{u}+2\zeta_3-\frac{\pi^2}{6}\right] - \frac{(\pi^2+48)}{6 \, (u-1)}\; , \\
h_{0}(u)  &=&-\frac{u (2u-1)\ln(u)}{(u-1)^2} + \frac{u}{u-1}\; , \\
h_{1}(u)  &=& \frac{u (2u-1)\ln^2(u)}{(u-1)^2}+\frac{u (2u-1)\PL{2}{1-u}}{(u-1)^2} -\frac{u (5u-4)\ln(u)}{(u-1)^2}+ \frac{2u}{u-1}\; , \\
h_{2}(u)  &=& \frac{u (2u-1)}{(u-1)^2} \left[\ln^2(u)\ln(1-u)-\frac{2}{3}\ln^3(u)-\frac{5\pi^2}{12}\ln(u)+\PL{3}{1-u}+2\PL{3}{u}\right.\nnb\\
           && -2  \zeta_3-\ln(u)\ln(1-u)+\ln^2(u)-2\ln(u)-\PL{2}{u}\Bigg] + \frac{\pi^2 \, u}{6\,(u-1)^2} \nnb\\	   
	  &&+\frac{3 u}{u-1} \left[\ln^2(u)-\ln(u)\ln(1-u)- \PL{2}{u}-2\ln(u)+\frac{11\pi^2}{36}+\frac{4}{3}\right]\; .
\eea

\subsection{Two-loop results}

The two-loop part of the form factor $F_1$ reads
\bea
F_1^{(2)} &=& C_F^2 \left\{\frac{1}{2\eps^4}+\frac{1}{\eps^3}\left[L+j_{-3}(u) \right] + \frac{1}{\eps^2}\left[L^2+2 L \, j_{-3}(u) + j_{-2}(u)\right]\right. \nnb\\
            &&              + \frac{1}{\eps}\left[\frac{2 L^3}{3}+2 L^2 \, j_{-3}(u) +2 L \, j_{-2}(u)+ j_{-1}(u)\right]\nnb\\
	    &&             \left. + \left[\frac{L^4}{3}+\frac{4}{3} L^3 \, j_{-3}(u) +2 L^2 \, j_{-2}(u)+2 L \,
	    j_{-1}(u)+j_{0}(u)\right]\right\}\nnb\\
            && + C_F \, C_A \left\{-\frac{11}{12\eps^3} +\frac{1}{\eps^2}\left[-\frac{11}{6} L+k_{-2}(u) \right]+\frac{1}{\eps}\left[-\frac{11}{6} L^2+2 L \, k_{-2}(u) +k_{-1}(u)\right] \right.\nnb\\
	    && \left. + \left[-\frac{11}{9} L^3+2 L^2 \, k_{-2}(u) +2 L \, k_{-1}(u)+ k_{0}(u)\right] \right\}\nnb\\
	    && + C_F \, t_f \, n_f \left\{ \frac{1}{3\eps^3} +\frac{1}{\eps^2}\left[\frac{2}{3} L+p_{-2}(u) \right]+\frac{1}{\eps}\left[\frac{2}{3} L^2+2 L \, p_{-2}(u) +p_{-1}(u)\right] \right.\nnb\\
	    && \left. + \left[\frac{4}{9} L^3+2 L^2 \, p_{-2}(u) +2 L \, p_{-1}(u)+ p_{0}(u)\right] \right\}\nnb\\
	    && + C_F \, t_f \left\{ \frac{4}{3\eps^3} +\frac{1}{\eps^2}\left[\frac{8}{3} L+q_{-2}(u) \right]+\frac{1}{\eps}\left[\frac{8}{3} L^2+2 L \, q_{-2}(u) +q_{-1}(u)\right] \right.\nnb\\
	    && \left. + \left[\frac{16}{9} L^3+2 L^2 \, q_{-2}(u) +2 L \, q_{-1}(u)+ q_{0}(u)\right] \right\}\label{eq:C1twoloop}
\eea
with
\bea
j_{-3}(u) &=& \frac{5}{2} - 2\ln(u) \; ,\\
j_{-2}(u) &=& 4 \ln^2(u)-2 \ln(u)\ln(1-u)-2 \PL{2}{u} - \frac{\ln(u)}{u-1}-8 \ln(u) +\frac{5\pi^2}{12}+\frac{73}{8} \; ,\\
j_{-1}(u) &=& -\frac{16}{3} \ln^3(u) +6 \ln^2(u)\ln(1-u) + \frac{3\ln^2(u)}{u-1}+14 \ln^2(u)- \frac{\ln(1-u)\ln(u)}{u-1} \nnb \\
          && -8 \ln(u)\ln(1-u)- \frac{13\ln(u)}{2\, (u-1)} -\frac{5\pi^2}{3} \ln(u) - \!\frac{55}{2} \ln(u)
	     +4 \ln(u)\PL{2}{u} - \!\frac{\PL{2}{u}}{u-1}\nnb \\
	   &&   -8 \PL{2}{u} + 2 \PL{3}{1-u}+ 4 \PL{3}{u} + \frac{\pi^2}{6 \, (u-1)} - \frac{31}{3} \, \zeta_3
	   +\frac{9\pi^2}{4} +\frac{213}{8}\; , \\
j_{0}(u)  &=& \frac{(3u^3-8u^2+52u-56)}{6 \,u^3} \Bigg[\ln^4(1-u)+2 \pi^2 \ln^2(1-u)-4 \ln^3(1-u)\ln(u) \nnb \\
          && \left. +24 \PL{4}{\frac{u}{u-1}}\right] +\frac{8 \, (5u^3+12u^2-78u+84)}{3\, u^3} \ln(u)\,\zeta_3 - \frac{28}{3} \ln^3(u)\ln(1-u)\nnb \\
	  && +\frac{(13u^3-12u^2+78u-84)}{u^3} \ln^2(u)\ln^2(1-u)  + \frac{5\ln(1-u)\ln^2(u)}{u-1} \nnb \\
	  &&+\frac{(30u^3-39u^2+56u-3)}{u^3} \ln^2(u)\ln(1-u) + \frac{10\pi^2}{3} \ln^2(u)- \frac{15\ln(1-u)\ln(u)}{2 \,
	  (u-1)}\nnb\\
	  &&+\frac{2 \, (5u^3-2u^2+13u-14)}{u^3} \left[{\rm Li}_2^2(u)+2 \ln(u)\ln(1-u)\PL{2}{u}\right]- \frac{14 \ln^3(u)}{3\,(u-1)}\nnb\\
	  && +\frac{(u^3+2u-2)}{u^3} \left[-16 \, {\rm HPL}(\{-2,2\},1-u)+\frac{4\pi^2}{3} \PL{2}{u-1}\right]+ \frac{25\ln^2(u)}{u}\nnb\\
	  && +\frac{8 \, (5u^2-10u+4)}{u^2} \left[ {\rm HPL}(\{-1,2\},1-u)+\frac{\pi^2}{12}\ln(2-u)\right]+ \frac{59}{2}\ln^2(u) \nnb\\
	  &&-\frac{8 }{u-1} \, {\rm HPL}(\{-1,2\},1-u)- \frac{2\pi^2\ln(2-u)}{3\,(u-1)}+ \frac{16}{3} \ln^4(u) - \frac{52}{3} \ln^3(u)\nnb\\
	  && -\frac{\pi^2 \, (17u^3-32u^2+192u-208)}{3\,u^3} \ln(1-u)\ln(u)-\frac{153}{2} \ln(u)+ \frac{25\ln^2(u)}{2\,(u-1)}\nnb\\
	  && +\frac{(19u^2-180u+100)}{2\,u^2} \left[\ln(1-u)\ln(u)+\PL{2}{u}\right] + \frac{\pi^2\ln(u)}{6\,(u-1)} - \frac{49\ln(u)}{2 \,(u-1)}\nnb\\
	  && +\frac{(u^2-3)}{u^3} \left[\ln^2(u)\ln(1+u)+\pi^2\ln(1+u)+2 \ln(u) \PL{2}{-u}-2 \PL{3}{-u}\right]\nnb\\
	  &&-\frac{8\pi^2 (u+3) (3u-4)}{3\,u^2} \ln(u)+\frac{2 \, (16u^3-51u^2+84u-3)}{u^3} \ln(u)\PL{2}{u}\nnb\\
	  && - \frac{15 \PL{2}{u}}{2 \, (u-1)}+ \frac{6 \ln(u)\PL{2}{u}}{u-1}-\frac{\pi^2 \, (11u^3-16u^2+88u-96)}{3\,u^3} \PL{2}{u}\nnb\\
	  && +\frac{4 \, (5u^3-8u^2+52u-56)}{u^3} \ln(u)\PL{3}{1-u} - 4 \ln^2(u)\PL{2}{u}- \frac{\PL{3}{1-u}}{u-1}\nnb \\
	  && -\frac{6 \, (2u^3-2u^2-4u-1)}{u^3} \PL{3}{1-u}- 8 \ln(u)\PL{3}{u}- \frac{2\PL{3}{u}}{u-1}\nnb \\
	  && -\frac{2 \, (2u^3-63u^2+112u-3)}{u^3} \PL{3}{u}-\frac{2 \, (9u^3-20u^2+114u-124)}{u^3} \PL{4}{1-u}\nnb \\
	  &&+\frac{4 \, (u^3-8u^2+52u-56)}{u^3} \PL{4}{u}+\frac{\pi^2 \, (21u^2+1168 u-1344)}{48 \,u^2}\nnb\\
	  &&+\frac{\pi^4 \, (2453u^3-960u^2+6096u-6576)}{2160 \, u^3}+\frac{2\, (14u^3-234u^2+336u-9) \zeta_3}{3 \, u^3}\nnb\\
	  &&-\frac{4 \, \pi^2 \,(u-4)\ln(2)}{u}+ \frac{5\pi^2}{4\,(u-1)}+ \frac{2\zeta_3}{u-1}+\frac{1327}{16}+2 \, c_0 \; ,\\
k_{-2}(u) &=& \frac{11}{3} \ln(u) +\frac{\pi^2}{12} - \frac{58}{9} \;,\\
k_{-1}(u) &=&-\frac{22}{3} \ln^2(u) + \frac{22}{3} \ln(u)\ln(1-u)+\frac{22}{3} \PL{2}{u} +\frac{11\ln(u)}{3 \, (u-1)} - \frac{\pi^2}{3} \ln(u) \nnb\\
	  &&+\frac{166}{9} \ln(u) +\frac{11}{2} \zeta_3-\frac{131\pi^2}{72}-\frac{6301}{216} \; ,\\
k_{0}(u)  &=& -\frac{(4u^2-10u+7)}{3 \,u^3} \Bigg[\ln^4(1-u)+2 \pi^2 \ln^2(1-u)-4 \ln^3(1-u)\ln(u) \nnb \\
          && \left. +24 \PL{4}{u} +24 \PL{4}{\frac{u}{u-1}}\right] -\frac{2 \, (7u^3-16u^2+40u-28)}{u^3} \ln(u)\,\zeta_3 \nnb\\
	  && -\frac{(4u^3+12u^2-30u+21)}{u^3} \ln^2(u)\ln^2(1-u) \nnb\\
	  &&-\frac{(106u^3+33u^2-84u-9)}{6 \,u^3} \ln^2(u)\ln(1-u) + \frac{2\pi^2}{3} \ln^2(u)+ \frac{22\ln(1-u)\ln(u)}{3 \,
	  (u-1)}\nnb\\
	  &&-\frac{(4u^3+4u^2-10u+7)}{u^3} \left[{\rm Li}_2^2(u)+2 \ln(u)\ln(1-u)\PL{2}{u}\right] \nnb\\
	  && +\frac{(u^3+2u-2)}{u^3} \left[8 \, {\rm HPL}(\{-2,2\},1-u)-\frac{2\pi^2}{3} \PL{2}{u-1}\right]+ \frac{17\ln^2(u)}{2 \,u}\nnb\\
	  && -\frac{4 \, (5u^2-10u+4)}{u^2} \left[ {\rm HPL}(\{-1,2\},1-u)+\frac{\pi^2}{12}\ln(2-u)\right]-\frac{547}{18}\ln^2(u) \nnb\\
	  &&+\frac{4}{u-1} \, {\rm HPL}(\{-1,2\},1-u)+ \frac{\pi^2\ln(2-u)}{3\,(u-1)}+ \frac{88}{9} \ln^3(u)\nnb\\ 
	  && +\frac{8 \pi^2 \, (4u^2-11u+8)}{3\,u^3} \ln(1-u)\ln(u)+\frac{4129}{54} \ln(u)- \frac{35\ln^2(u)}{6\,(u-1)}\nnb\\
	  && +\frac{(215u^2-108u+153)}{9\,u^2} \left[\ln(1-u)\ln(u)+\PL{2}{u}\right] - \frac{2\pi^2\ln(u)}{3\,(u-1)} + \frac{533\ln(u)}{18 \,(u-1)}\nnb\\
	  && -\frac{(u^2-3)}{2 \, u^3} \left[\ln^2(u)\ln(1+u)+\pi^2\ln(1+u)+2 \ln(u) \PL{2}{-u}-2 \PL{3}{-u}\right]\nnb\\
	  &&+\frac{\pi^2 (155 u^2-180 u+216)}{18\,u^2} \ln(u)-\frac{3 \, (2u^3+8u^2-14u-1)}{u^3} \ln(u)\PL{2}{u}\nnb\\
	  && + \frac{22 \PL{2}{u}}{3 \, (u-1)}+\frac{4 \pi^2 \, (2u-3)^2}{3\,u^3} \PL{2}{u}-\frac{8 \, (u^3+4u^2-10u+7)}{u^3} \ln(u)\PL{3}{1-u}\nnb\\
	  && -\frac{(2u^3+93u^2-90u+9)}{3 \, u^3} \PL{3}{1-u} -\frac{(70u^3-111u^2+168u+9)}{3 \, u^3} \PL{3}{u}\nnb \\ 
	  && +\frac{2 \, (20u^2-58u+43)}{u^3} \PL{4}{1-u}-\frac{\pi^2 \, (230u^2-405 u+378)}{27 \,u^2}\nnb\\
	  &&-\frac{\pi^4 \, (269u^3+480u^2-1236u+876)}{1080 \, u^3}+\frac{(461u^3-396u^2+1008u+54) \zeta_3}{18 \, u^3}\nnb\\
	  &&+\frac{2 \, \pi^2 \,(u-4)\ln(2)}{u}- \frac{11\pi^2}{9\,(u-1)}-\frac{146461}{1296}- c_0 \; ,\\ 
p_{-2}(u) &=& \frac{20}{9} - \frac{4}{3} \ln(u) \;,\\
p_{-1}(u) &=&\frac{8}{3} \ln^2(u) - \frac{8}{3} \ln(u)\ln(1-u)-\frac{8}{3} \PL{2}{u} -\frac{4\ln(u)}{3 \, (u-1)} -\frac{56}{9} \ln(u) +\frac{13\pi^2}{18}+\frac{557}{54} \; ,\nnb \\
          && \\
p_{0}(u)  &=& -\frac{32}{9} \ln^3(u)+\frac{16}{3} \ln^2(u)\ln(1-u)+ \frac{8\ln^2(u)}{3\,(u-1)}+ \frac{112\ln^2(u)}{9}- \frac{8\ln(u)\ln(1-u)}{3\,(u-1)}\nnb \\
          && - \frac{112}{9}\ln(u)\ln(1-u)-\frac{86\ln(u)}{9\,(u-1)}-\frac{26\pi^2}{9} \ln(u)-\frac{706}{27} \ln(u)- \frac{8\PL{2}{u}}{3\,(u-1)}-\frac{74}{9}\zeta_3\nnb \\
	  &&- \frac{112}{9} \PL{2}{u}+\frac{16}{3} \PL{3}{1-u}+\frac{32}{3} \PL{3}{u}+\frac{4\pi^2}{9\,(u-1)}+\frac{106\pi^2}{27}+\frac{11813}{324}\; ,\\
q_{-2}(u) &=& \frac{10}{3} - \frac{8}{3} \ln(u) \;,\\
q_{-1}(u) &=&\frac{8}{3} \ln^2(u) - \frac{8}{3} \ln(u)\ln(1-u)-\frac{8}{3} \PL{2}{u} -\frac{4\ln(u)}{3 \, (u-1)} -4 \ln(u) +\frac{2\pi^2}{3}+8 \; , \\
q_{0}(u)  &=& -\frac{16}{9} \ln^3(u)+\frac{8}{3} \ln^2(u)\ln(1-u)+ \frac{4\ln^2(u)}{3\,(u-1)}+ 4\ln^2(u)- \frac{8\ln(u)\ln(1-u)}{3\,(u-1)}\nnb \\
          && +\frac{8 \, (5u^3-39u^2+54u-16)}{9 \, u^3} \left[\ln(u)\ln(1-u)+\PL{2}{u}\right]-\frac{86\ln(u)}{9\,(u-1)}-\frac{4\pi^2}{3} \ln(u)\nnb\\
	  && -\frac{2 \, (409u^2-660u+192)}{27\,u^2} \ln(u)- \frac{8\PL{2}{u}}{3\,(u-1)}+\frac{16\, (u^3-3 u+3)}{3 \, u^3} \PL{3}{1-u}\nnb \\
	  &&+\frac{16}{3} \PL{3}{u}+\frac{4\pi^2}{9\,(u-1)}-\frac{\pi^2(11u^2+12u-24)}{9\,u^2}-\frac{16\, (5u^3-9u+9)}{9\, u^3}\zeta_3\nnb\\
	  && +\frac{10543u^2-9144 u+2304}{162\,u^2}\; .
\eea
We comment on the r\^ole of the constant $c_0$ in the Appendix. The two-loop part of the form factor $F_2$ reads
\bea
F_2^{(2)} &=& C_F^2 \left\{\frac{1}{\eps^2} \, r_{-2}(u) + \frac{1}{\eps}\left[2 L \, r_{-2}(u) + r_{-1}(u)\right] + \left[2 L^2 \, r_{-2}(u)+2 L \, r_{-1}(u)+r_{0}(u)\right]\right\}\nnb\\
            && + C_F \, C_A \left\{\frac{1}{\eps}\, s_{-1}(u) +\left[2 L \, s_{-1}(u)+s_{0}(u) \right] \right\}\nnb\\
	    && + C_F \, t_f \, n_f \left\{ \frac{1}{\eps} \, t_{-1}(u)  +\left[2 L \, t_{-1}(u)+t_{0}(u) \right]    \right\}\nnb\\
	    && + C_F \, t_f \left\{ \frac{1}{\eps} \, v_{-1}(u)  +\left[2 L \, v_{-1}(u)+v_{0}(u) \right]  \right\}\label{eq:C2twoloop}
\eea
with
\bea
r_{-2}(u) &=& - g_{0}(u) \; ,\\
r_{-1}(u) &=& \frac{2 \, u}{(u-1)^2} \left[\PL{2}{1-u} + 3 \ln^2(u) \right] - \frac{(11u-4)\ln(u)}{(u-1)^2}+\frac{9}{u-1}\; , \\
r_{0}(u)  &=& \frac{8 \; (2u-7)}{3 \,u^3} \Bigg[\ln^4(1-u)+2 \pi^2 \ln^2(1-u)-4 \ln^3(1-u)\ln(u) \nnb \\
          &&  -24\ln(u)\,\zeta_3+9\ln^2(u)\ln^2(1-u)+3 \, {\rm Li}_2^2(u)+6 \ln(u)\ln(1-u)\PL{2}{u} \nnb\\
	  && \left. +24\ln(u)\PL{3}{1-u}+24 \PL{4}{u}+24 \PL{4}{\frac{u}{u-1}}\right]+ \frac{50\ln^2(u)}{u}\nnb\\
	  &&+\frac{2 \, (11u^2+52u-3)}{u^3} \ln^2(u)\ln(1-u) + \frac{(23u-36)}{(u-1)^2}\left[\ln(1-u)\ln(u)+\PL{2}{u}\right]\nnb\\
	  && +\frac{8}{u^3} \left[8 \, {\rm HPL}(\{-2,2\},1-u)-\frac{2\pi^2}{3} \PL{2}{u-1}\right] - \frac{2 (3u-8)}{(u-1)^2}\ln^2(u)\ln(1-u) \nnb\\
	  && +\frac{32 \, (u+2)}{u^2} \left[ {\rm HPL}(\{-1,2\},1-u)+\frac{\pi^2}{12}\ln(2-u)\right]-\frac{(25u^2-49u+28)}{(u-1)^3}\ln^2(u) \nnb\\
	  &&-\frac{16\,u}{(u-1)^2} \left[ {\rm HPL}(\{-1,2\},1-u)+\frac{\pi^2}{12}\ln(2-u)+\frac{7}{12} \ln^3(u)\right]\nnb\\ 
	  && -\frac{32 \pi^2 \, (4u-13)}{3\,u^3} \ln(1-u)\ln(u) -\frac{4\, (11u-25)}{u^2} \left[\ln(1-u)\ln(u)+\PL{2}{u}\right] \nnb\\
	  && -\frac{2\,(u^2+\!4u+\!3)}{u^3} \!\left[\ln^2(u)\ln(1+u)+\pi^2\ln(1+u)+\!2 \ln(u) \PL{2}{-u}-2 \!\PL{3}{-u}\right]\nnb\\
	  &&+\frac{32\pi^2 (u+6)}{3\,u^2} \ln(u) +\!\frac{4 (17u^2+80u-3)}{u^3} \ln(u)\PL{2}{u}  - \!\frac{12 (3u-4)}{(u-1)^2} \ln(u)\PL{2}{u} \nnb\\
	  &&-\frac{64 \pi^2 \, (u-3)}{3\,u^3} \PL{2}{u}+\frac{4\,(8u+3)}{u^3} \PL{3}{1-u}-\frac{2\,(9u-8)}{(u-1)^2} \PL{3}{1-u} \nnb\\
	  &&  -\frac{4\,(23u^2+108u-3)}{u^3} \PL{3}{u} +\frac{4\,(15u-16)}{(u-1)^2} \PL{3}{u}-\frac{16 \, (10u-31)}{u^3} \PL{4}{1-u}\nnb \\ 
	  && +\frac{2\pi^2 \, (25 u-84)}{3 \,u^2}-\frac{\pi^2 \, (41 u-54)}{6 \,(u-1)^2}+\frac{2 \pi^4 \, (40u-137)}{45 \, u^3}- \frac{\pi^2(31u-32)}{3\,(u-1)^2}\ln(u)\nnb\\
	  &&- \frac{8\,(3u-1)}{(u-1)^2}\ln(u)+\frac{4\,(22u^2+116u-3) \zeta_3}{u^3}-\frac{4\,(15u-16)}{(u-1)^2} \zeta_3+\frac{31}{u-1} \; ,\\
s_{-1}(u) &=& \frac{22u\ln(u)}{3 \, (u-1)^2}-\frac{22}{3 \, (u-1)} \; ,\\
s_{0}(u)  &=& \frac{2 \; (4u-7)}{3 \,u^3} \Bigg[\ln^4(1-u)+2 \pi^2 \ln^2(1-u)-4 \ln^3(1-u)\ln(u) \nnb \\
          &&  -24\ln(u)\,\zeta_3+9\ln^2(u)\ln^2(1-u)+3 \, {\rm Li}_2^2(u)+6 \ln(u)\ln(1-u)\PL{2}{u} \nnb\\
	  && \left. +24\ln(u)\PL{3}{1-u}+24 \PL{4}{u}+24 \PL{4}{\frac{u}{u-1}}\right]+ \frac{17\ln^2(u)}{u}\nnb\\
	  &&-\frac{(u^2-32u-3)}{u^3} \ln^2(u)\ln(1-u) + \frac{4 \, (17u-6)}{3 \,(u-1)^2} \left[\ln(1-u)\ln(u)+\PL{2}{u}\right]\nnb\\
	  && -\frac{4}{u^3} \left[8 \, {\rm HPL}(\{-2,2\},1-u)-\frac{2\pi^2}{3} \PL{2}{u-1}\right] - \frac{4}{u-1}\ln^2(u)\ln(1-u) \nnb\\
	  && -\frac{16 \, (u+2)}{u^2} \left[ {\rm HPL}(\{-1,2\},1-u)+\frac{\pi^2}{12}\ln(2-u)\right]-\frac{5\,(13 u-6)}{3 \,(u-1)^2}\ln^2(u) \nnb\\
	  &&+\frac{8\,u}{(u-1)^2} \left[ {\rm HPL}(\{-1,2\},1-u)+\frac{\pi^2}{12}\ln(2-u)\right]\nnb\\ 
	  && -\frac{64 \pi^2 \, (u-2)}{3\,u^3} \ln(1-u)\ln(u) -\frac{2\, (16u-17)}{u^2} \left[\ln(1-u)\ln(u)+\PL{2}{u}\right] \nnb\\
	  && +\frac{(u^2+4u+3)}{u^3} \!\left[\ln^2(u)\ln(1+u)+\pi^2\ln(1+u)+2 \ln(u) \PL{2}{-u}-2 \!\PL{3}{-u}\right]\nnb\\
	  &&+\frac{4\pi^2 (u+18)}{3\,u^2} \ln(u) -\frac{2 (2u^2-46u-3)}{u^3} \ln(u)\PL{2}{u}  - \frac{12}{u-1} \ln(u)\PL{2}{u} \nnb\\
	  &&-\frac{8 \pi^2 \, (4u-9)}{3\,u^3} \PL{2}{u}+\frac{2\,(5u^2+34u-3)}{u^3} \PL{3}{1-u}-\frac{4\PL{3}{1-u}}{u-1}  \nnb\\
	  && + \frac{6\,(u^2-20u-1)}{u^3} \PL{3}{u} +\frac{16 \PL{3}{u}}{u-1} -\frac{4 \, (20u-43)}{u^3} \PL{4}{1-u}\nnb \\ 
	  && +\frac{2\pi^2 \, (3 u-14)}{u^2}-\frac{2\pi^2 \, (8u+3)}{9 \,(u-1)^2}+\frac{\pi^4 \, (40u-73)}{45 \, u^3}- \frac{4\pi^2(3u-2)}{3\,(u-1)^2}\ln(u)\nnb\\
	  &&+\frac{(323u-54)}{9\,(u-1)^2}\ln(u)-\frac{2\,(2u^2-52u-3) \zeta_3}{u^3}-\frac{16\, \zeta_3}{u-1} -\frac{401}{9\,(u-1)} \; ,\\
t_{-1}(u) &=& - \frac{8u\ln(u)}{3 \, (u-1)^2}+\frac{8}{3 \, (u-1)} \; ,\\
t_{0}(u)  &=& \frac{4\,u}{9\,(u-1)^2} \left[12\ln^2(u)+12\PL{2}{1-u}-19\ln(u)\right]+\frac{124}{9 \, (u-1)} \; ,\\
v_{-1}(u) &=& t_{-1}(u)  \; ,\\
v_{0}(u)  &=& \frac{8\,u}{3\,(u-1)^2} \left[\ln^2(u)+2\PL{2}{1-u}\right]-\frac{4 (79 u-60)}{9\,(u-1)^2}\ln(u)+\frac{80\ln(u)}{3\,u}\nnb\\
          && +\frac{8\,(u+2)}{3\,u^2}\left[2\ln(u)\ln(1-u)+2\PL{2}{u}+\pi^2\right]+\frac{124}{9 \, (u-1)} \nnb\\
	  && +\frac{32\PL{3}{1-u}}{u^3}-\frac{32\zeta_3}{u^3}-\frac{104}{3\,u}\; .
\eea
Finally, the two-loop part of the form factor $F_3$ reads
\bea
F_3^{(2)} &=& C_F^2 \left\{\frac{1}{\eps^2} \, w_{-2}(u) + \frac{1}{\eps}\left[2 L \, w_{-2}(u) + w_{-1}(u)\right] + \left[2 L^2 \, w_{-2}(u)+2 L \, w_{-1}(u)+w_{0}(u)\right]\right\}\nnb\\
            && + C_F \, C_A \left\{\frac{1}{\eps}\, x_{-1}(u) +\left[2 L \, x_{-1}(u)+x_{0}(u) \right] \right\}\nnb\\
	    && + C_F \, t_f \, n_f \left\{ \frac{1}{\eps} \, y_{-1}(u)  +\left[2 L \, y_{-1}(u)+y_{0}(u) \right]    \right\}\nnb\\
	    && + C_F \, t_f \left\{ \frac{1}{\eps} \, z_{-1}(u)  +\left[2 L \, z_{-1}(u)+z_{0}(u) \right]  \right\}\label{eq:C3twoloop}
\eea
with
\bea
w_{-2}(u) &=& - h_{0}(u) \; ,\\
w_{-1}(u)  &=& -\frac{u (2u-1)}{(u-1)^2}\left[3\ln^2(u)+\PL{2}{1-u}\right]+\frac{u (24u-17)}{2\,(u-1)^2}\ln(u) - \frac{9u}{2\,(u-1)}\; , \\
w_{0}(u)  &=& \frac{4 \; (2u^2-16u+21)}{3 \,u^3} \Bigg[\ln^4(1-u)+2 \pi^2 \ln^2(1-u)-4 \ln^3(1-u)\ln(u) \nnb \\
          &&  -24\ln(u)\,\zeta_3+9\ln^2(u)\ln^2(1-u)+3 \, {\rm Li}_2^2(u)+6 \ln(u)\ln(1-u)\PL{2}{u} \nnb\\
	  && \left. +24\ln(u)\PL{3}{1-u}+24 \PL{4}{u}+24 \PL{4}{\frac{u}{u-1}}\right]- \frac{75\ln^2(u)}{u}-\frac{31u}{2\,(u-1)}\nnb\\
	  &&-\frac{(16u^3-47u^2+160u-9)}{u^3} \ln^2(u)\ln(1-u) + \frac{6 (u-2)}{(u-1)^2} \ln(u)\PL{2}{u}\nnb\\
	  &&+ \frac{(u+12)}{2 \,(u-1)^2}\left[\ln(1-u)\ln(u)+\PL{2}{u}\right] - \frac{(7u-2)}{(u-1)^2}\ln^2(u)\ln(1-u) + \frac{28\ln^3(u)}{3}\nnb\\
	  && +\frac{32(2u-3)}{u^3} \left[{\rm HPL}(\{-2,2\},1-u)-\frac{\pi^2}{12} \PL{2}{u-1}\right] -\frac{2\,(13u-14)}{(u-1)^2} \PL{3}{u}\nnb\\
	  && -\frac{16 \, (u^2-u+6)}{u^2} \left[ {\rm HPL}(\{-1,2\},1-u)+\frac{\pi^2}{12}\ln(2-u)\right]+ \frac{\pi^2(29u-\!30)}{6\,(u-1)^2}\ln(u)\nnb\\
	  &&+\frac{8\,(3u-2)}{(u-1)^2} \left[ {\rm HPL}(\{-1,2\},1-u)+\frac{\pi^2}{12}\ln(2-u)+\frac{7}{12} \ln^3(u)\right]+8\pi^2\ln(2)\nnb\\
	  && -\frac{(24 u^3-73u^2+69u-24)}{2\,(u-1)^3}\ln^2(u) -\frac{16 \pi^2 \, (4u^2-30u+39)}{3\,u^3} \ln(1-u)\ln(u)\nnb \\
	  &&  -\frac{2\, (6u^2-71u+75)}{u^2} \left[\ln(1-u)\ln(u)+\PL{2}{u}\right] -\frac{\pi^2 (u^2-80u+288)}{3\,u^2} \ln(u) \nnb\\
	  && +\frac{(2u^3+3u^2+8u+9)}{u^3} \!\left[\ln^2(u)\ln(1+u)+\pi^2\ln(1+u)+2 \ln(u) \PL{2}{-u}\right.\nnb\\
	  && \left. -2 \!\PL{3}{-u}\right]-\frac{2 (10u^3-69u^2+244u-\!9)}{u^3} \ln(u)\PL{2}{u}+\frac{(11u-\!10)}{(u-1)^2} \PL{3}{1-u}\nnb\\
	  &&-\frac{32 \pi^2 \, (u^2-7u+9)}{3\,u^3} \PL{2}{u}+\frac{2\,(3u^3+8 u^2-28 u-9)}{u^3} \PL{3}{1-u} \nnb\\
	  &&  +\frac{2\,(4u^3-91u^2+328u-9)}{u^3} \PL{3}{u}-\frac{8 \, (10u^2-72u+93)}{u^3} \PL{4}{1-u}\nnb \\ 
	  && +\frac{\pi^2 \, (5u^2-390 u+504)}{6 \,u^2}+\frac{\pi^2 \, (17 u-30)}{12 \,(u-1)^2}-\frac{2\,(14u^3-94u^2+344u-9) \zeta_3}{u^3}\nnb\\
	  &&+ \frac{(79u^2-67u+4)}{2\,(u-1)^2}\ln(u)+\frac{\pi^4 \, (40u^2-314u+411)}{45 \, u^3}+\frac{2\,(13u-14)}{(u-1)^2} \zeta_3 \; ,\\
x_{-1}(u)  &=& -\frac{11u (2u-1)}{3 \, (u-1)^2}\ln(u)+ \frac{11u}{3\,(u-1)}\; , \\
x_{0}(u)  &=& \frac{(8u^2-24u+21)}{3 \,u^3} \Bigg[\ln^4(1-u)+2 \pi^2 \ln^2(1-u)-4 \ln^3(1-u)\ln(u) \nnb \\
          &&  -24\ln(u)\,\zeta_3+9\ln^2(u)\ln^2(1-u)+3 \, {\rm Li}_2^2(u)+6 \ln(u)\ln(1-u)\PL{2}{u} \nnb\\
	  && \left. +24\ln(u)\PL{3}{1-u}+24 \PL{4}{u}+24 \PL{4}{\frac{u}{u-1}}\right]- \frac{51\ln^2(u)}{2\,u}+\frac{347u+54}{18\,(u-1)}\nnb\\
	  &&-\frac{(2u^3-51u^2+92u+9)}{2\,u^3} \ln^2(u)\ln(1-u) + \frac{6}{u-1} \ln(u)\PL{2}{u}\nnb\\
	  &&- \frac{2(39u-28)}{3 \,(u-1)^2}\left[\ln(1-u)\ln(u)+\PL{2}{u}\right] + \frac{2}{u-1}\ln^2(u)\ln(1-u) \nnb\\
	  && -\frac{16(2u-3)}{u^3} \left[{\rm HPL}(\{-2,2\},1-u)-\frac{\pi^2}{12} \PL{2}{u-1}\right] -\frac{8}{u-1} \PL{3}{u}\nnb\\
	  && +\frac{8 \, (u^2-u+6)}{u^2} \left[ {\rm HPL}(\{-1,2\},1-u)+\frac{\pi^2}{12}\ln(2-u)\right]+ \frac{2\pi^2(5u-4)}{3\,(u-1)^2}\ln(u)\nnb\\
	  &&-\frac{4\,(3u-2)}{(u-1)^2} \left[ {\rm HPL}(\{-1,2\},1-u)+\frac{\pi^2}{12}\ln(2-u)\right]-4\pi^2\ln(2)\nnb\\
	  && +\frac{(94u^2-53u-6)}{6\,(u-1)^2}\ln^2(u) -\frac{16 \pi^2 \, (4u^2-13u+12)}{3\,u^3} \ln(1-u)\ln(u)\nnb \\
	  &&  -\frac{(56u^2-216u+153)}{3\,u^2} \left[\ln(1-u)\ln(u)+\PL{2}{u}\right] -\frac{2\pi^2 (2u^2-29u+54)}{3\,u^2} \ln(u) \nnb\\
	  && -\frac{(2u^3+3u^2+8u+9)}{2 \, u^3} \!\left[\ln^2(u)\ln(1+u)+\pi^2\ln(1+u)+2 \ln(u) \PL{2}{-u}\right.\nnb\\
	  && \left. -2 \!\PL{3}{-u}\right]-\frac{(8u^3-78u^2+134u+9)}{u^3} \ln(u)\PL{2}{u}+\frac{2\PL{3}{1-u}}{u-1} \nnb\\
	  &&-\frac{4 \pi^2 \, (8u^2-28u+27)}{3\,u^3} \PL{2}{u}-\frac{(6u^3-41 u^2+98 u-9)}{u^3} \PL{3}{1-u} \nnb\\
	  &&  +\frac{(14u^3-105u^2+176u+9)}{u^3} \PL{3}{u}-\frac{2 \, (40u^2-136u+129)}{u^3} \PL{4}{1-u}\nnb \\ 
	  && +\frac{\pi^2 \, (76u^2-285 u+378)}{9 \,u^2}+\frac{\pi^2 \, (30 u-19)}{9 \,(u-1)^2}-\frac{(4u^3-102u^2+160u+9) \zeta_3}{u^3}\nnb\\
	  &&- \frac{(760u^2-329u-162)}{18\,(u-1)^2}\ln(u)+\frac{\pi^4 \, (80u^2-246u+219)}{90 \, u^3}+\frac{8\zeta_3}{u-1}  \; ,\\
y_{-1}(u)  &=& \frac{4u (2u-1)}{3 \, (u-1)^2}\ln(u)- \frac{4u}{3\,(u-1)}\; , \\
y_{0}(u)  &=& -\frac{8\,u(2u-1)}{3\,(u-1)^2} \left[\ln^2(u)+\PL{2}{1-u}\right]+\frac{2 \,u(68 u-49)}{9 \, (u-1)^2}\ln(u)-\frac{62u}{9 \, (u-1)} \; ,\\
z_{-1}(u) &=& y_{-1}(u)  \; ,\\
z_{0}(u)  &=& -\frac{4\,u (2u-1)}{3\,(u-1)^2}\ln^2(u)-\frac{8\, (3u-2)}{3\,(u-1)^2}\PL{2}{1-u}+\frac{2 (147 u-128)}{9\,(u-1)^2}\ln(u)\nnb\\
          && +\frac{8 (17 u-93)}{9\,u}\ln(u)+\frac{8\,(9u-22)}{3\,u^2}\left[\ln(u)\ln(1-u)+\PL{2}{u}\right]-\frac{16 (2u-3)}{u^3}\zeta_3\nnb\\
	  && -\frac{2 (55 u-24)}{9 \, (u-1)} -\frac{4\pi^2(2 u^2-3 u+18)}{9\, u^2}+\frac{16 (2 u-3)}{u^3}\PL{3}{1-u}+\frac{284}{3\,u} \; .
\eea
As already mentioned in the introduction, our results have been compared 
analytically with those of~\cite{Bonciani:2008wf} and complete agreement 
has been obtained.
\section{Numerical evaluation and 
conclusion}
\label{sec:conclude}

\begin{figure}[p]
\begin{center}
  \includegraphics[width=0.6\textwidth]{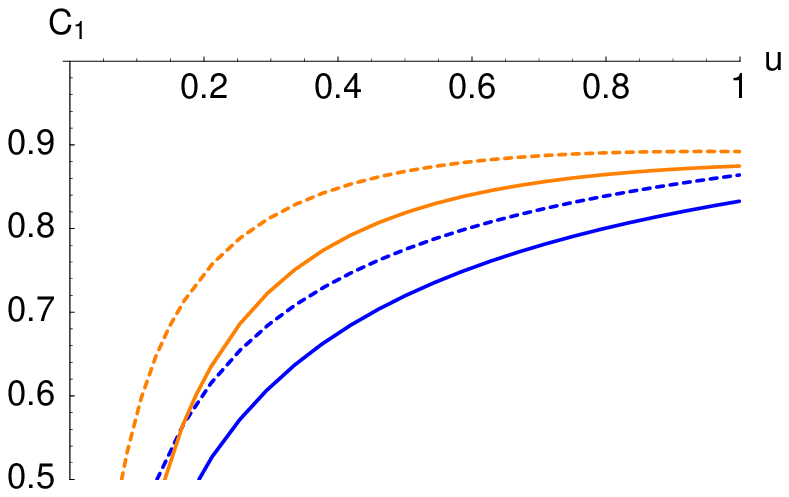}\\
  \includegraphics[width=0.63\textwidth]{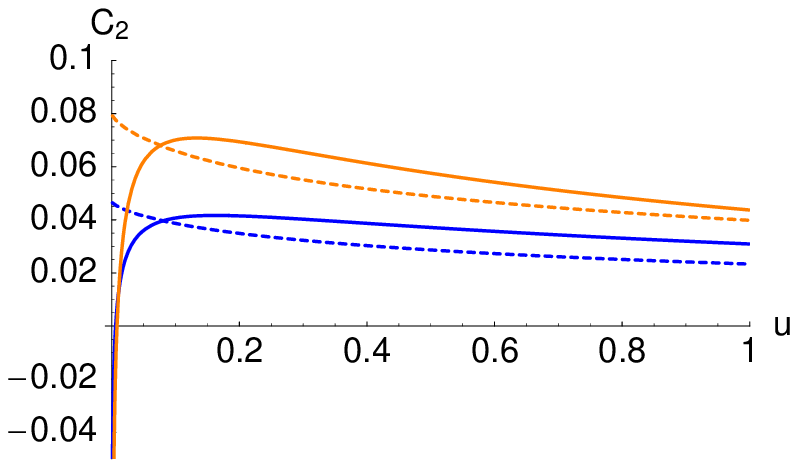}\\
  \includegraphics[width=0.63\textwidth]{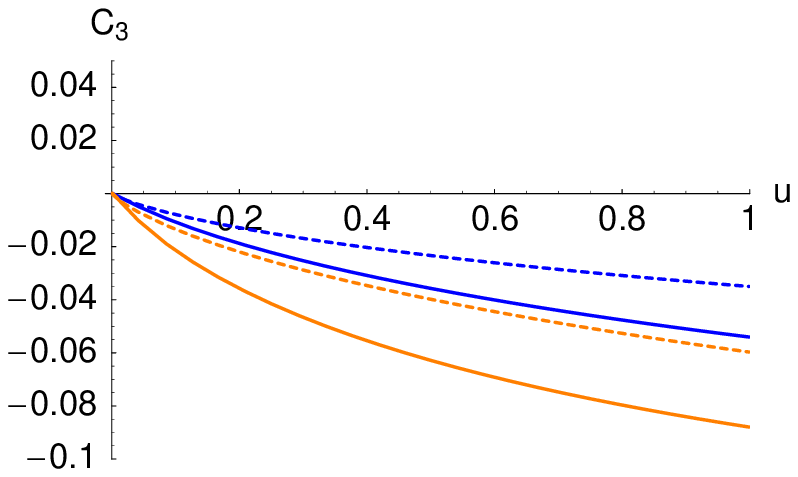}\\
\caption{The matching coefficients $C_i(u)$ 
($i=1,2,3$ from top to bottom) in the one-loop (dashed) and 
two-loop (solid) approximation. The blue/dark grey curves 
refer to $\mu=m_b=4.8\,$GeV, the orange/light grey curves 
to $\mu=1.5\,$GeV.}
\label{fig:Cs}
\end{center}
\end{figure}

In Fig.~\ref{fig:Cs} we show the coefficient functions $C_i(u)$ in 
the one- (dashed) and two-loop (solid) approximation. For this purpose, 
we define the renormalized matching coefficients in the $\overline{\rm MS}$ 
scheme by subtracting minimally the infrared poles. More precisely, 
from $C_i = Z_J^{-1} F_i$, we obtain
\be
C_i = C_i^{(0)} + \frac{{\alpha_s^{(4)}}}{4\pi} \, C_i^{(1)}+ \frac{{\alpha_s^{(4)}}^2}{(4\pi)^2} \, C_i^{(2)}+ {\cal O}(\as^3) \, ,
\ee
with
\begin{eqnarray}
C_i^{(0)} & = & F_i^{(0)},\\
C_i^{(1)} & = & F_i^{(1)} +[Z_J^{-1}]^{(1)} \,F_i^{(0)},\\
C_i^{(2)} & = & F_i^{(2)} + F_i^{(1)} \left\{ \,[Z_J^{-1}]^{(1)} + \xi_{45}^{(1)} + Z_{\alpha}^{(1)}\right\} +
[Z_J^{-1}]^{(2)} \,F_i^{(0)} \, .
\label{cms}
\end{eqnarray} 
Expressions for the $F_i^{(j)}$ have been given in earlier sections. $[Z_J^{-1}]^{(1)}$ can be derived by inverting~(\ref{eq:ZJ}) and
taking the coefficient of ${\alpha_s^{(4)}}/(4\pi)$. $\xi_{45}^{(1)}$ is the coefficient of ${\alpha_s^{(4)}}/(4\pi)$
in~(\ref{eq:asD}), and $Z_{\alpha}^{(1)}$ the one of ${\alpha_s}/(4\pi)$ in~(\ref{zalpha}).
$[Z_J^{-1}]^{(2)}$ is defined to subtract the remaining 
pole parts in $ F_1^{(2)} + F_1^{(1)} \{ \,[Z_J^{-1}]^{(1)} + \xi_{45}^{(1)} + Z_{\alpha}^{(1)}\}$, 
such that $C_1^{(2)}$ is IR-finite. Note that to evaluate 
(\ref{cms}) one needs the ${\cal O}(\epsilon^2)$ terms of 
$F_i^{(1)}$ given in Sec.~\ref{subsec:oneloop}, since $[Z_J^{-1}]^{(1)}$ 
contains a $1/\epsilon^2$ singularity.

These 
coefficients depend on the IR factorization scale, which cancels 
only in the product of hard, jet- and shape-function factors. 
In physical applications the factorization scale ranges between 
$\sqrt{m_b \Lambda_{\rm QCD}}$ and $m_b$. To illustrate the size of 
the new two-loop correction, we therefore evaluate 
$C_i(u)$ at $\mu=m_b=4.8\,$GeV (blue/dark grey curves) and 
at $\mu=1.5\,$GeV (orange/light-grey curves). We emphasize that the 
difference between these two choices is not a theoretical error -- it is 
compensated by a corresponding scale dependence of the convolution 
$J\star S$. The value of the strong coupling is 
$\alpha^{(5)}_s(m_b)=\alpha^{(4)}_s(m_b)=0.22$, and $\alpha^{(4)}_s(1.5\,\mbox{GeV})=0.3753$. 
Comparison of the dashed and solid curves of the same colour 
in Fig.~\ref{fig:Cs} shows that the two-loop corrections are generally 
very moderate, if not small, except in the region of small $u$, 
where increasing powers of large logarithms take over. The implications 
of this result for the $|V_{ub}|$ determination remain to be 
investigated. The impact of the ${\cal O}(\alpha_s^2)$ terms 
depends on the combination $C_i C_j \,J\star S$, and the numerical size 
of the two-loop correction to the jet and shape function has not 
yet been analyzed. A straightforward evaluation of the partonic 
structure functions $W_i$ in the shape-function region 
indicates sizeable two-loop effects. A 
reanalysis of existing $B\to X_u\ell \bar \nu$ 
decay distribution data with 
${\cal O}(\alpha_s^2)$ accuracy taking into account renormalization group 
summation and a model of the $B$ meson shape function is therefore 
well motivated.

\vspace*{0.5em}
\noindent
\subsubsection*{Acknowledgement}
We thank W.~Bernreuther for useful discussions, and R.~Bonciani and 
A.~Ferroglia for performing a comparison of
Eqs.~(\ref{eq:C1twoloop}),~(\ref{eq:C2twoloop}), and~(\ref{eq:C3twoloop}).
This work is supported in part by the 
DFG Sonder\-forschungsbereich/Transregio~9 
``Computergest\"utzte Theoretische Teilchenphysik'' 
and the Swiss National Science Foundation (SNF). 
X.-Q.~Li acknowledges support from the Alexander-von-Humboldt Stiftung.
M.B. acknowledges hospitality from the University of Z\"urich and the CERN
theory group, where part of this work was performed.

\appendix

\section{Master integrals}
\label{ap:masters}
\begin{figure}[t]
\centering
\includegraphics[width=0.9\textwidth]{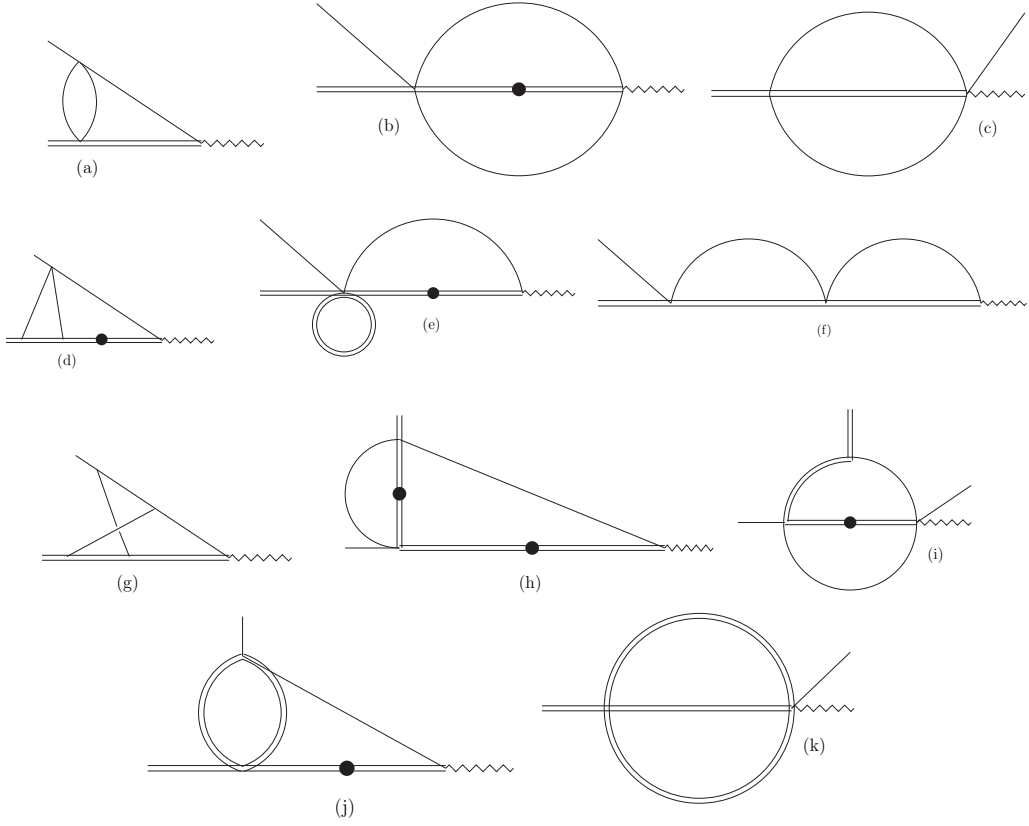}
\caption{Two-loop master integrals needed for the calculation. Double lines are massive, and single lines are massless. All diagrams
stand for scalar integrals with unit numerator. Dots on lines represent squared propagators. Topologies with one or more dots stand for
the undotted diagram and all diagrams with {\textit{one}} single dot, i.e.\ topology (h) stands for three diagrams.\label{fig:masters}}
\end{figure}
The two-loop calculation gives rise to 18 master integrals, which are depicted in Fig.~\ref{fig:masters}. All necessary master integrals
were already computed in~\cite{Bell:2006tz}, 
which chooses a slightly different basis compared to the present work.
Certain individual master integrals can also be found in
\cite{Bonciani:2008az,Fleischer:1999hp,Argeri:2002wz,Fleischer:1998nb,Aglietti:2003yc}.
We find almost perfect agreement with the results in~\cite{Bell:2006tz}. In the case of the crossed six-line master integral $I_g$ in
Fig.~\ref{fig:masters}(g) we improve the numerical accuracy of the boundary condition of the finite part. We therefore give our results
for $I_g$ explicitly. Our integration measure reads
\be
\int \! \left[dk\right] \equiv \loopint{k} \; .
\ee
We define the prefactor
\be
\ESGamma \equiv \frac{1}{\left(4\pi\right)^{D/2} \Gamma(1-\eps)}\;,
\ee
and express our results in terms of
\be
x \equiv \frac{(p_b-p)^2+i\eta}{m_b^2} = 1 - u \; .
\ee
Here $+i\eta$ stems from the $+i\eta$ prescription which we tacitly assume to be included in the propagators of the integral below. We
find
\bea
I_g &=& \int \! \left[dk_1\right] \!\int \! \left[dk_2\right] \frac{1}{\left[(k_2+p_b)^2-m_b^2\right] (k_2+p)^2
\left[(k_1+p_b)^2-m_b^2\right] (k_2-k_1+p)^2} \nnb\\
&& \hspace*{74pt}\times \frac{1}{(k_2-k_1)^2 \, k_1^2} \nnb \\
&=& - \ESGamma^2 \, (m_b^2)^{-2-2\eps}
\left\{\frac{c_g^{(-4)}}{\eps^4}+\frac{c_g^{(-3)}}{\eps^3}+\frac{c_g^{(-2)}}{\eps^2}+\frac{c_g^{(-1)}}{\eps}+c_g^{(0)}\right\}
\eea
with
\bea
c_g^{(-4)} &=&\frac{1}{12 \,(1-x)^2} \; , \nnb\\
c_g^{(-3)} &=&-\frac{\ln(1-x)}{3 \,(1-x)^2} \; , \nnb\\
c_g^{(-2)} &=&\frac{1}{72 \,(1-x)^2} \left[48 \ln^2(1-x) - 5\pi^2\right] \; ,\nnb\\
c_g^{(-1)} &=&\frac{1}{36 \,(1-x)^2} \left[-32\ln^3(1-x)+10 \pi^2 \ln(1-x) - 267\,\zeta_3 \right] \; ,\nnb\\
c_g^{(0)} &=&\frac{1}{(1-x)^2} \left[\frac{8}{9} \ln^4(1-x) - \frac{5}{9}\pi^2\ln^2(1-x)+8 \ln(1-x)\PL{3}{x}\right. \nnb\\
            && \hspace*{50pt} \left. +\frac{65}{3} \ln(1-x) \,\zeta_3 + 4 \, {\rm Li}_2^2(x) + c_0\right] \; .
\eea
The constant $c_0$, obtained with the package 
MB.m~\cite{Czakon:2005rk} from a three-dimensional Mellin-Barnes representation, equals $-60.2493267(10)$, 
where the number in parenthesis gives the uncertainty of the last two digits displayed. The number excludes $c_0 = -89\pi^4/144$ which
was found in~\cite{Bell:2006tz}\footnote{In the revised version of~\cite{Bell:2006tz} the constant was replaced by a numerical value on
which we agree within the given error bars.}. It suggests $c_0 = -167\pi^4/270$, in
agreement with the proposal in~\cite{Bonciani:2008wf}. To date there does, 
however, not exist a value for $c_0$ that is derived completely  
analytically. Except for $c_0$, we obtained all other terms in 
the master integrals by analytical steps, that is, without fitting 
rational numbers to numerical values.

\end{document}